# ALLOWING FOR EQUAL OPPORTUNITIES FOR ARTISTS IN MUSIC RECOMMENDATION: A POSITION PAPER

**Christine Bauer**
Johannes Kepler University Linz
Institute of Computational Perception
`christine.bauer@jku.at`

## ABSTRACT

Promoting diversity in the music sector is widely discussed on the media. While the major problem may lie deep in our society, music information retrieval contributes to promoting diversity or may create unequal opportunities for artists. For example, considering the known problem of popularity bias in music recommendation, it is important to investigate whether the short head of popular music artists and the long tail of less popular ones show similar patterns of diversity—in terms of, for example, age, gender, or ethnic origin—or the popularity bias amplifies a positive or negative effect.

I advocate for reasonable opportunities for artists—for (currently) popular artists and artists in the long-tail alike—in music recommender systems. In this work, I represent the position that we need to develop a deep understanding of the biases and inequalities because it is the essential basis to design approaches for music recommendation that provide reasonable opportunities. Thus, research needs to investigate the various reasons that hinder equal opportunity and diversity in music recommendation.

## 1. INTRODUCTION

Creating and maintaining diversity is an important and widely discussed topic in our society [27]. Thereby the debates on diversity are dominated by the challenges in promoting diversity as our society is prone to lay ground for unequal opportunities with respect to, for example, age, disability, gender, ethnic origin, religion, or sexual orientation throughout our society.

The issue of unequal opportunities is also relevant and a highly topical subject in the music sector. Some people voiced their concerns that there is a general discrimination of female artists [3, 16, 20, 24]. A similar inequality problem exists with respect to the little representation of black artists (especially black female artists) in high-popularity playlists on online music platforms [19,20].

While the major problem may lie far beneath online music platforms or the music sector at large, the vast possibilities of music information retrieval and recommendation may contribute tremendously in promoting diversity, inclusion, and equity—but may also be used to (intentionally or unintentionally) create unreasonable imbalances.

For instance, it is widely known that algorithms used for music recommendation are frequently prone to popularity bias [12]. This is a burden to inclusion, as such algorithms prioritize popular items and almost disregard the long tail of less popular items. In other words, the spectrum of suggested items is limited to a proportionally small set of items. As popularity bias is a common phenomenon in algorithmic filtering, research came up with diversity measures [15, 23] and there are various attempts to introduce diversity to recommendation algorithms [5,6]. Studies (e.g., [9]) have shown that an increase in diversity has a positive effect on user experience, while the ideal degree of diversity may depend on user characteristics [10,13,21].

I postulate that we need to develop a deep understanding of the biases and inequalities because it is the essential basis to design approaches for music recommendation that are free from undesired biases and inequalities.

When we take a human-centric approach to music information retrieval (MIR), we need to consider all kinds of roles involved in MIR—not just the user. In this work, I put the—previously neglected—artists' perspective in the loop. With the goal to provide reasonable opportunities for artists—for (currently) popular artists and artists in the long-tail alike—in music recommender systems, I take the position that research needs to investigate the various reasons that hinder equal opportunity and diversity in music recommendation.

This position paper is structured as follows: Section 2 presents the complexity of bias in music recommendation. Section 3 puts the artists' perspective into the loop. Section 4 presents the fundamental research questions that have to be addressed to allow for equal opportunity for artists and promoting diversity in music consumption.

## 2. THE COMPLEXITY OF BIAS

Music recommendation relies on algorithmic decision-making. And an emerging body of literature has shown that algorithmic decision-making can go wrong in multiple ways [25], due to algorithmic problems, data sparsity, or actors gaming the system (e.g., via click manipulation). Typical problems include popularity bias, cold start prob-lem, shilling attacks, grey-sheep problem, synonymy, as well as scalability and latency problems [14]. This leads to severe problems for society—from filter bubbles [18] to the reproduction and amplification of stereotypes and discrimination [22] to cognitive bias and humans' overconfidence in algorithmic results [11]. Addressing these problems, there is a growing body of literature on fairness, accountability, and transparency in machine learning and artificial intelligence [2, 7, 17].

Still, while some aspects of bias in data and algorithms are subject of interest in research and draw attention on the media (e.g., filter bubble and popularity bias), other biases are not addressed or may even not have been identified yet.

## 3. THE NEGLECTED ARTIST IN MUSIC INFORMATION RETRIEVAL

In the music information retrieval (MIR) community (and related communities), research on diversity typically takes the perspective of the system—for instance, to mitigate the cold-start problem [4]—or the user (here: music consumer)—to better meet user preferences [13, 26]. The perspective of the item suppliers is considered only occasionally. For instance, Reference [1] raise awareness that recommender systems in multi-stakeholder environments may be fair for one stakeholder while being unfair for other stakeholders. Reference [8] proposes an approach with the goal to provide all artists in a collection with the opportunity of being listened in recommendations. Taking a human-centric approach to MIR systems, the goal is to include the artists' perspective in MIR research.

## 4. RESEARCH DIRECTIONS

Taking a human-centric perspective with the aim to allow for equal opportunity for artists and promoting diversity in music consumption, requires to address fundamental research questions concerning potential bias in current systems and, generally, in music consumption. For instance:

**Research Question 1.** *How is diversity in terms of, for example, age, disability, gender, ethnic origin, religion, or sexual orientation of artists represented in the long tail of the popularity distribution?*
*How is diversity represented in the short head of popular artists?*
*How does the diversity in the long tail and the short head relate to each other, and to the entire population?*

**Research Question 2.** *How does the popularity of music items reflect inherent user taste?*
*How is the popularity of music items affected by what is offered on online music platforms, on playlist, in recommendations, in advertising, etc.?*

Understanding bias is a prerequisite to address its various facets and mitigate them. One concrete research question could be formulated as follows:

**Research Question 3.** *What is the influence of using timbre of the singing voice for music recommendation on the artist gender distribution in recommended items?*
*If recommendations allow for little diversity in timbre, items will likely be sung by same-gender singers.*

Overall, the goal of future work is to investigate the various facets reasons that hinder equal opportunity and diversity in music recommendation. A deep understanding of the biases and inequalities is the essential basis to design approaches for music recommendation that provide reasonable opportunities for artists—for (currently) popular artists and artists in the long-tail alike.

## 5. ACKNOWLEDGMENTS

This research is supported by the Austrian Science Fund (FWF): V579.